%% file: main.tex
\pdfoutput=1
\documentclass[a4paper]{article}
\usepackage{INTERSPEECH2020}
\usepackage{amsmath,graphicx}

\usepackage{hyperref}
\input{MACPdef}

\usepackage{color}
\usepackage{colortbl}
\definecolor{light-gray}{gray}{0.8}

\title{Data Techniques for Online End-to-end Speech Recognition}
\name{Yang Chen$^{\star}$ \thanks{Work done while Yang was a summer intern at Amazon Alexa.} \qquad Weiran Wang$^{\dagger}$ \thanks{Work done while Weiran Wang was at Amazon Alexa.} \qquad  I-Fan Chen$^{\dagger\dagger}$ \qquad  Chao Wang$^{\dagger\dagger}$}
\address{$^{\star}$Georgia Institute of Technology $\qquad$ $^{\dagger}$Salesforce Research $\qquad$ $^{\dagger\dagger}$Amazon Alexa}
\email{ychen3411@gatech.edu \hspace{0.5em} weiran.wang@salesforce.com \hspace{0.5em} \{ifanchen,wngcha\}@amazon.com}

\begin{document}
\maketitle

\begin{abstract}
Practitioners often need to build ASR systems for new use cases in a short amount of time, given limited in-domain data. While recently developed end-to-end methods largely simplify the modeling pipelines, they still suffer from the data sparsity issue. 
In this work, we explore a few simple-to-implement techniques for building online ASR systems in an end-to-end fashion, with a small amount of transcribed data in the target domain. These techniques include data augmentation in the target domain, domain adaptation using models previously trained on a large  source domain, and knowledge distillation on non-transcribed  target domain data, using an adapted bi-directional model as the teacher; they are applicable in real scenarios with different types of resources. Our experiments demonstrate that each technique is independently useful in the  improvement of the online ASR performance in the target domain.
\end{abstract}

\input{intro}
\input{augmentation}
\input{adaptation}
\input{distillation}

\input{conclusion}

\section{Acknowledgements}

The authors would like to thank Hao Tang and Qingming Tang for
discussions on WSJ baselines, Harsha Sundar for issues regarding
signal processing, Bowen Shi for suggesting the self-training 
experiment, and the Amazon ASR team for sharing source domain data.
Weiran Wang is grateful to Prof. Karen Livescu and Hao Tang, from whom he
learned ASR.

%\cleardoublepage

\bibliographystyle{IEEEtran}
\bibliography{refs}
\end{document}

%% file: MACPdef.tex
\newcommand{\caja}[4][1]{{%
    \renewcommand{\arraystretch}{#1}%
    \begin{tabular}[#2]{@{}#3@{}}%
      #4%
    \end{tabular}%
    }}

{%
\begin{list}{#1}{
\vspace{-\topsep}
\vspace{-\partopsep}
\setlength{\itemindent}{0cm}
\setlength{\rightmargin}{0cm}
\setlength{\listparindent}{0cm}
\settowidth{\labelwidth}{#1}
\setlength{\leftmargin}{\labelwidth}
\addtolength{\leftmargin}{\labelsep}
\setlength{\itemsep}{0cm}
}%
}%
{%
\end{list}
\vspace{-\topsep}
\vspace{-\partopsep}
}

{\begin{enumerate}%
}%
{\end{enumerate}}

%% file: intro.tex
\section{Introduction}
\label{s:intro}
% \vspace*{-2ex}

End-to-end speech recognition systems have gained increasing popularity, due to the simplicity of their modeling pipelines (without initial alignment for warm start), easiness of deployment (with more light-weight decoders), and comparable performance to the state-of-the-art. The two major classes of end-to-end models are variants of connectionist temporal classification (CTC) ~\cite{Graves_06a,Miao_15a,Collob_16a,Graves_12a}, and attention-based sequence to sequence models~\cite{Chorow_15a,Chan_16a,Watanab_17a,Zeyer_18a}.
The advantages of end-to-end methods make them viable choices for developing ASR systems from scratch in a short amount of time. On the other hand, since these methods try to avoid expert knowledge and to learn everything from data, 
they may suffer more from the data sparsity issue than traditional methods.
% as can be seen from their performance on standard benchmarks.

In this work, we are concerned with building end-to-end ASR systems for new use cases given limited in-domain data, which is a task frequently faced by practitioners in the early stage of the development. Furthermore, we are interested in deploying models with low latency, which rules out bi-directional architectures (despite that they give superior performance~\cite{Miao_16a}). Since it is less trivial to deploy attention-based models in an online fashion~\cite{Jaitly_18a,Chiu_18a},  and inspired by the success of ASR systems on mobile devices~\cite{He_18a},  
we mainly consider unidirectional architecture and CTC-based models. More recent variants of CTC such as AutoSeg~\cite{Collob_16a,Pratap_18a} and RNN-transducer~\cite{Rao_17a,He_18a} are left for future investigation.

\noindent\textbf{Contributions} In this work, we explore a few existing and new easy-to-implement techniques in this scenario, depending on the availability of different types of resources: 
\begin{itemize}
\item data augmentation techniques in the target domain~\cite{Ko_15a,Park_19a},
\item domain adaptation from an initial model trained in a source domain with large amount of data~\cite{Neto_95a,Yao_12a},
\item a new teacher-student learning framework if non-transcribed data is available in the target domain~\cite{Mosner_19a,Chen_19b,Hari_19a,distill,distill2,Huang_18a,guide_ctc1}.
\end{itemize}
We found each technique to be independently useful and we obtain significant accuracy improvement by combining them.
% all of them. 

The common intuition behind our techniques is about data: augmenting target domain data, leveraging source domain data, and  generating pseudo-labels so we have more (noisy) training data. This work shows how much we can gain by carefully manipulating data, even using relatively simple acoustic models.
We believe our setting is quite common and our findings are useful to practitioners.

In the following, we first demonstrate the abovementioned techniques on the Wall Street Journal (WSJ) corpus,\footnote{Obtained from LDC under the catalog numbers LDC93S6B and LDC94S13B.} which yields about 50\% relative improvement over a baseline, when trained on 15 hours of supervised in-domain data.
After that, we provide experimental results on an in-house dataset of conversational speech. Our methods behave consistently for this more challenging dataset as for WSJ, and we achieve 25.4\% relative improvement in WER over a baseline that is pretrained on 15000 hours out-of-domain data.

\section{Setup on WSJ}
\label{s:setup}

For most of the paper, we use the the WSJ corpus as our target domain. In particular, unless mentioned otherwise, we use the \emph{si84} partition (7040 utterances) as the transcribed target domain data, while the \emph{si284} partition (37.3K utterances) is used as unsupervised target domain data. We use the \emph{dev93} partition (503 utterances) as development set for hyper-parameter tuning, and the \emph{eval92} partition (333 utterances) as test set. 
We report both phone error rate (PER) and word error rate (WER) on the evaluation sets. 
The units used by our CTC acoustic models are the 351 position-dependent phones together with the \texttt{<blank>} symbol. Acoustic model training is done with Tensorflow~\cite{Abadi_15a}, and we use its beam search algorithm for phone-level decoding with a beam size of $20$.
For word-level decoding, we use the WFST-based framework~\cite{Miao_15a} with the lexicon and trigram language model (with a 20K vocabulary size) provided by the kaldi \emph{s5} recipe~\cite{Povey_11a}, and run beam search with a beam size of $20$ on per-frame log-likelihood produced by the acoustic model. 
The conversion from posterior (acoustic model output) to likelihood uses a uniform prior on non-blank symbols, and a different prior for \texttt{<blank>} which is tuned on the dev set~\cite{Sak_15a}.  
\texttt{<NOISE>} and \texttt{<UNK>} are removed both from ground truth and decoding results for calculating WER. 

Our source domain CTC acoustic models are trained on a 15000 hour subset of data used for training the Amazon Transcribe system~\cite{Amazon_18a}. We use two models trained in the source domain, one consisting of 5 uni-directional LSTM layers~\cite{HochreitSchmid97a} of 512 units for domain adaptation, and the other consisting of 5 bi-directional LSTM layers of 512 units in each direction, as the teacher model for knowledge distillation on non-transcribed target data. The source domain CTC models are trained with a different set of units than those used for the target domain. We use 5 LSTM layers instead of a shallower one because we find that, with careful tuning, the deeper architecture performs better even trained on \emph{si84} (e.g., a 3 uni-directional LSTM architecture gives 25.6\% PER on \emph{dev93}, versus the 23.71\% obtained by 5 uni-directional LSTMs shown in Table~\ref{t:result_augmentation}).

For both source and target domains, 40 dimensional LFBEs are extracted with a window size of 25ms and hop size of 10ms. We perform per-speaker mean normalization on the LFBEs in the target domain. Furthermore, we stack every 3 consecutive frames to reduce input sequence length by three times to speeds up training and decoding, where the initial frame index for stacking is randomly selected from $\{0,1,2\}$ during training and fixed to $0$ during evaluation; this already provides a form of data augmentation~\cite{Sak_15a}.

For acoustic model training, we use the ADAM optimizer~\cite{KingmaBa15a} with minibatches of $8$ utterances, and an initial learning rate searched over the grid $\{$0.0002, 0.0005, 0.001, 0.002$\}$ when the model is initialized randomly, and searched over the grid $\{$0.00002, 0.00005, 0.0001, 0.0002$\}$ when adapting the source domain models (the smaller learning rates are important for domain adaptation). For all model training, we apply dropout~\cite{Srivas_14a} with rate tuned over $\{$0.0, 0.1, 0.2, 0.4$\}$, which is effective with small amount of training data. Each model is trained up to $50$ epochs and the iteration which yields the best PER is selected for evaluation.

%% file: augmentation.tex
\section{Data augmentation}
\label{s:augmentation}
% \vspace*{-2ex}

A natural approach for alleviating the data sparsity issue in the target
domain is to augment the supervised training set with different perturbed
versions~\cite{Ko_15a,Zhou_17a,Park_19a}. Here we mainly explore speed perturbation and spectral masking.

\begin{figure}[t]
    \centering
    \includegraphics[width=0.95\linewidth,bb=120 70 890 630,clip]{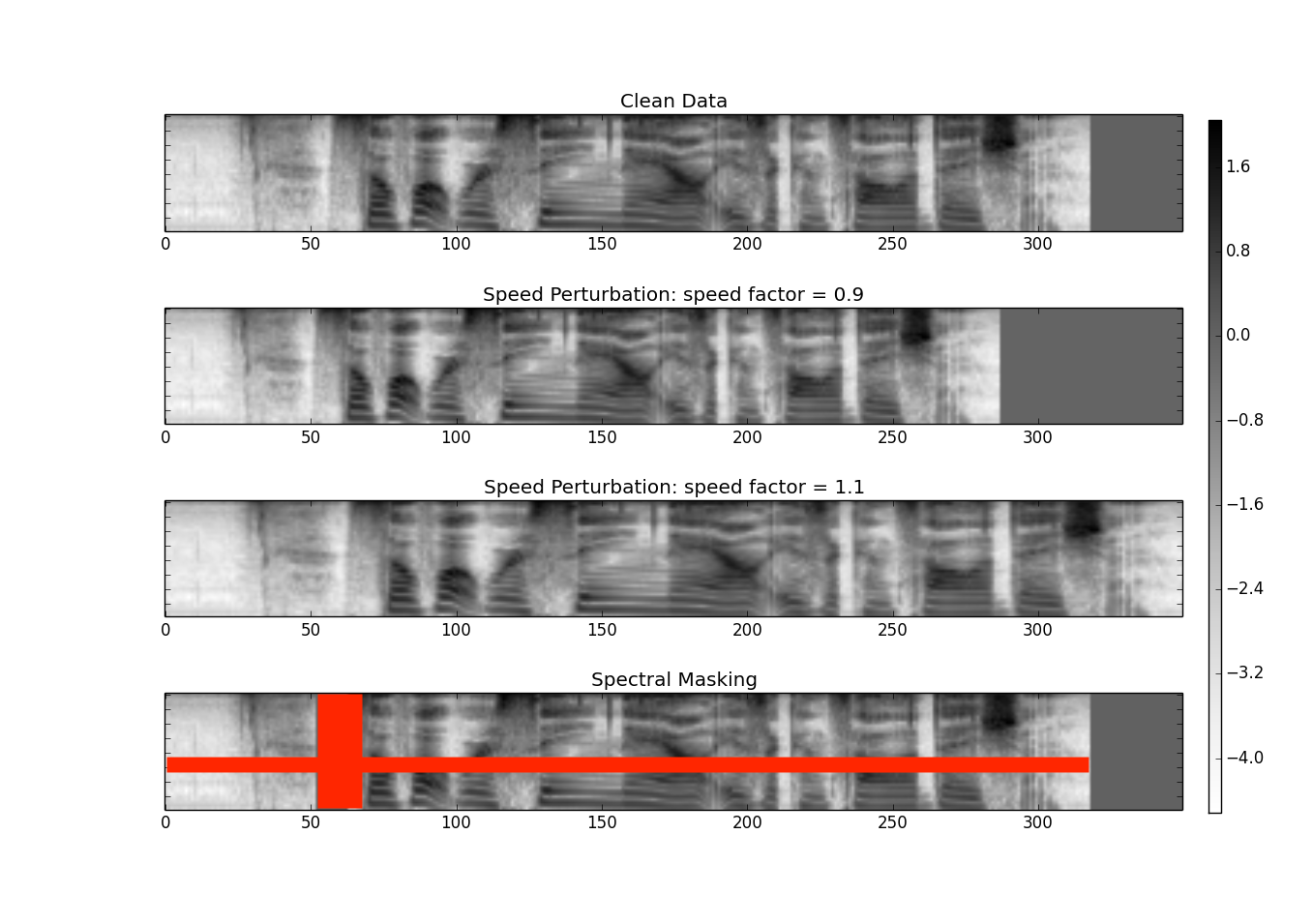}
    \caption{Illustration of data augmentation techniques.}
    \label{figure_augmentation}
\vspace*{-1ex}
\end{figure}

\subsection{Speed Perturbation}
% \vspace*{-1.5ex}

Speed perturbation is an augmentation technique that
produces a warped time signal, and it is shown to improve performance
on LVCSR~\cite{Ko_15a}. The implementation we use is a mix of that
of~\cite{Ko_15a} and the time warping method of~\cite{Park_19a}: instead
of perturbing the audio~\cite{Ko_15a}, we apply
linear interpolation on the spectrogram via the function
\texttt{interpolate.inter2d} from the \texttt{scipy} package to modify
temporal resolution. In other words, we treat the spectrogram as an image
and resize it in the time axis. 
The speed factors we use are $\{0.9, 1.0, 1.1\}$ as suggested in~\cite{Ko_15a}.
We adopt this implementation because it is easy to implement and we can
generate the perturbed versions on the fly, without the need of additional feature extraction from audio.

% \vspace*{-1ex}
\subsection{Spectral Masking}
% \vspace*{-1.5ex}

Recently,~\cite{Park_19a} proposed a set of data augmentation methods that operate on
the spectrogram directly, including time warping,
frequency masking, and time masking. They are shown to yield
significant improvement for the attention-based % sequence to sequence
model LAS~\cite{Chan_16a}, and we explore the two masking
techniques here within the CTC framework. 
The masking method applies zero masks to randomly selected
consecutive mel frequency channels and consecutive time steps on the
spectrogram. 
The two main parameters are F and T, corresponding to the
maximum length of frequency mask and time mask. We perform small grid
search for them, and set F to 8 and T to 16
in all of our experiments. Instead of applying masks to input features and
fixed it for training, the masks are generated and applied on the fly. We set the
probability of applying masks to a given input utterance to 0.5, for the
purpose of also exploiting clean data during training.
Figure~\ref{figure_augmentation} gives an illustration of the
abovementioned augmentation techniques.  Augmentation is performed before
stacking frames.

\begin{table}[t]
\centering
\caption{Performance (measured by PER in \%) of data augmentation techniques.}
\label{t:result_augmentation}
\begin{tabular}{lrr}
\hline
Model &\emph{dev93} &\emph{eval92}\\
\hline
Train on target & 23.71 & 18.32\\
\hline
+ Speed Perturbation & 21.67 & \\
+ Spectral Masking & 22.77 & \\
+ Both & \bf{20.55} &\bf{16.14} \\
\hline
\end{tabular}
\end{table}

Table~\ref{t:result_augmentation}
shows the result of speed perturbation, spectral masking, and the
combination of the two methods. Each technique helps to improve the performance
independently and the gains from both are additive. 
Although it was suggested by~\cite{Park_19a} that time warping is
not a major contributor to performance improvement, our observations on speed perturbation differ.
We have not explored reverberation and adding noisy to the audio, as the WSJ data is relatively clean.

%% file: adaptation.tex
\section{Domain adaptation}
\label{s:adaptation}
% \vspace*{-2ex}

In speech, the domain differences come from the acoustic condition, speaker, and style of speech (conversational vs. read, etc.). 
In addition to the mismatch in the inputs, there might be a mismatch in the output labels (number of classes, phones vs. characters, etc) between the source and target domains. 
In this section, we make use of an initial model that is well-trained on
a large amount of source domain data (15000 hours of conversational speech, see Section~\ref{s:setup}) which is denoted as Pretrain, and try to
transfer the modeling power to the target domain by finetuning from the
pre-trained model which is denoted as Finetune. In our case, since the source
domain model is trained with a different set of units, we replace the original softmax layer with a new randomly initialized one.

Apart from a good initialization, we could also insert an additional
linear layer between the target domain input and the pre-trained model
which is shared across all time steps, and only finetune this layer and
the softmax in the early stage of training, so the smaller number of
parameters in these adaptation layers are well trained with a small amount
of labeled data. 
This method is known as linear input networks
(LIN)~\cite{Neto_95a,Yao_12a}, and in this work we explore the method in the
context of end-to-end ASR. This linear layer is initialized as an identity mapping for the purpose of fast convergence.
For the first $10$ epochs over our training data, we fix the
weights of the 5 LSTM layers, and only update the LIN layer and the softmax layer. 

Table~\ref{t:result_adaptation} shows the results for domain adaptation,
with or without LIN. Pretraining + finetuning alone improves the
best performance from the previous section (training only on the target domain data) by
about 7\%  PER in absolute on the dev set, demonstrating that the modeling
power is well transferred to the target domain. On top of that, LIN and
data augmentation lead to consistent, further improvement. 
% \yang{Note we can not show the performance of Pretrain only since the pre-trained model has a newly initialized softmax layer for the target domain.}

\begin{table}[t]
\centering
\caption{Performance (PER in \%) of domain adaptation techniques. DataAug: Speed
  perturbation + Spectral masking. ``Train on Target + DataAug'' is taken from Table~\ref{t:result_augmentation}.}
\label{t:result_adaptation}
\begin{tabular}{lrr}
\hline
Model & \emph{dev93} & \emph{eval92} \\
\hline
Train on Target + DataAug & 20.55 & 16.14\\
\hline
Pretrain + Finetune & 13.60 & \\
$\qquad \qquad \qquad$ + DataAug & 13.01 &\\
\hline
Pretrain + Finetune + LIN & 10.72 &\\
$\qquad \qquad \qquad$ + DataAug &\bf{10.09} & \bf{7.32}\\
\hline
\end{tabular}
\end{table}

%% file: distillation.tex
\section{Distillation}
\label{s:distillation}
% \vspace*{-2ex}

In practice, it is often the case that the speed of data collection is much faster than the speed of annotation. Therefore, in the early stage of the development, we would have a small 
amount of transcribed speech and at the same time a relatively large amount of non-transcribed data. 
In this section, we generate pseudo-labels using a more powerful, bi-directional system on the unsupervised data and then distill the knowledge to an uni-directional system. 
Note that the bi-directional system is not deployable in the online mode; they are used as teachers on the unsupervised data to provide guidance for the uni-directional student model.

Let us connect our methodology to existing work. The pseudo-label approach is considered as sequence-level knowledge distillation by~\cite{distill}, in which the authors use a teacher model to generate top-k hypothesis on \emph{supervised} data and use a weighted sum of CTC loss on the top-k hypothesis for training the student. In contrast, we explore sequence-level knowledge distillation on the \emph{unsupervised} data, and incorporate a language model for label generation which leads to further improvement.
In addition to sequence-level distillation, frame-level distillation has been explored~\cite{distill, distill2, Huang_18a,guide_ctc1}.  In general, frame level distillation from bi-directional teacher to uni-directional student with per-frame KL divergence loss seems to degrade the ASR performance~\cite{distill,guide_ctc1}. 
This phenomenon is attributed to the different timing behavior between posteriors produced by bi-directional systems and uni-directional systems, and thus~\cite{guide_ctc1} proposes to align the CTC spike timing between bi-directional model and uni-directional model with a pre-trained guiding CTC model. On the other hand, \cite{Huang_18a} perform distillation possibly on unsupervised data, where both the teacher and the student are uni-directional models (with different depths), and propose to use pseudo-labels computed with a forward-backward algorithm as frame targets. In comparison to these work, our sequence level distillation approach on unsupervised data is different, and arguably simpler.

Our bi-directional teacher is initially trained in the same source domain, and then adapted in the same manner described in the previous section. And as expected, it achieves much better performance---5.36\% PER on \emph{dev93}---than the best uni-directional model so far.  
We consider two approaches to generate pseudo-labels at the token (phone) level on unsupervised data. The first is to use phone level beam search decoding result (with a beam size of 20). The second approach is to first apply WFST-based word decoder to obtain the most probable word sequence, and then convert it back to the phone sequence using the lexicon.  Here the language model used in word decoder is obtained from the source domain, to avoid using language model trained on \emph{si284}. This decoding approach yields a dev PER of 3.68\%. Clearly, incorporating language model can improve the quality of pseudo-labels. 

The uni-directional student model is then trained on both supervised utterances with ground truth and unsupervised utterances with the pseudo-labels, using the CTC objective. A discount factor is applied to the pseudo-label loss term, which is tuned by grid search. It turns out a discount of $1.0$ works best, perhaps because our pseudo-labels are relatively clean. For each gradient update, we use $8$ supervised utterances and $32$ unsupervised utterances. 
We provide in Table~\ref{t:result_distillation} the result of our trained student model, with the two types of pseudo-labels. To avoid cluttering, we only give the performance with adaptation and data augmentation; results without adaptation (not listed here) show similar behavior. We observe that knowledge distillation effectively explores the teacher's modeling power, and yields significant PER reduction.

To see how much improvement comes from the use of more (unsupervised) data versus the use of a more powerful teacher, we also perform semi-supervised learning with the self-training method of~\cite{Chen_19a}. The uni-directional model produces pseudo-labels using greedy beam search decoding on the fly for an unlabeled utterance, and the pseudo-labels are used by the augmented version of the same utterance (with speed perturbation and spectral masking) for CTC training. We train the uni-directional model with a discount factor set to $1.0$. We observe that while self-training does give sizable PER reduction, its performance is still inferior to that of the student guided by the bi-directional teacher.

\begin{table}[t]
\centering
\caption{Performance (measured by PER in \%) of knowledge distillation using unsupervised data. ``Pretrain + Finetune + LIN + DataAug'' is taken from Table~\ref{t:result_adaptation}.}
\label{t:result_distillation}
\begin{tabular}{lrr}
\hline
Model&\emph{dev93}&\emph{eval92}\\
\hline
Teacher (5 Bi-LSTM layers, Pretrain&& \\
+ Finetune + LIN + DataAug) & 5.36 &   \\
$\qquad \qquad \qquad \qquad$ + word decode & 3.68 & \\
\hline
Pretrain + Finetune + LIN + DataAug & 10.09 & 7.32 \\
\hline
+ Unsup KD (phone decode) & 8.37 &\\
+ Unsup KD (word decode)  & \textbf{8.25} & \textbf{6.04} \\
+ Self-training                     & 9.50 & \\
\hline
\end{tabular}
\end{table}

%% file: conclusion.tex
\section{Summary on WSJ}
\label{s:conclusion}
% \vspace*{-2ex}

\begin{table}[t]
\centering
\caption{ASR performance (measured by WER in \%) of previous work and our
  methods on \emph{eval92}, under similar settings. }
\label{t:result_wer}
\begin{tabular}{lr}
\hline
Model & WER \\
\hline
\cite{Baskar_19a} (attention, train on \emph{si84}, & \\ 
$\qquad\qquad\qquad$ unsup on \emph{si284} by ASR+TTS) & 20.30 \\
EESEN CTC~\cite{Miao_15a} (bi-directional, train on \emph{si284}) & 7.87\\
RNN-CTC~\cite{GravesJaitly14a} (bi-directional, train on \emph{si84}) &  13.50 \\
\hline
CTC (our uni-directional, train on \emph{si284}) & 11.98 \\
 \hline
This work (uni-directional, target \emph{si84}) & \\
Train on target (Sec.~\ref{s:augmentation}) & 17.72 \\
%Train on target + DataAug (Sec.~\ref{s:augmentation}) & 19.35 \\
%Adaptation + DataAug  (Sec.~\ref{s:adaptation}) &  12.33 \\
Adaptation + DataAug + KD on \emph{si284}  (Sec.~\ref{s:distillation}) & \textbf{7.58} \\
\hline
\end{tabular}
\end{table}

In Table~\ref{t:result_wer} we give the
WER results of our methods on \emph{eval92}, for the baseline and the best
model from % sections~\ref{s:augmentation}, \ref{s:adaptation} and
Section~\ref{s:distillation}.
To put our results in context, we also include in the table two CTC models from the
literature~\cite{Miao_15a,GravesJaitly14a} under similar settings in
terms of data and training objective.
The recent work~\cite{Baskar_19a} which uses similar data partition for
semi-supervised learning with attention model is also included.
We also train our uni-directional architecture on \emph{si284} to obtain another baseline, which
gives 11.98\% WER on \emph{eval92}. 
By combining the proposed techniques, we achieve more than 50\% relative
improvement over training only on the target domain data (17.72\%$\rightarrow$7.58\%), and
obtain an online system whose performance is better than that of a bi-directional system
trained on the same unsupervised data (13.50\% from~\cite{GravesJaitly14a}), and that of a
uni-directional system trained on more supervised target domain data (11.98\%).

\section{Experiments on conversational speech}
\label{s:iemocap}

We now demonstrate our approach on a more challenging in-house dataset. 
In particular, this dataset consists of IEMOCAP~\cite{iemocap} % a subset of MSP~\cite{msp}, 
and human-collected conversational speech. The supervised target domain data contains 10.9k/2.1k/2k utterances in the train/dev/test partitions, with disjoint set of speakers. In addition, we use 60 hours of non-transcribed data for unsupervised knowledge distillation. The token set we use for this dataset is the same as what we use for pretraining on source domain data.

\begin{table}[t]
\centering
\caption{ASR performance (measured by TER and WER in \%) of our
  methods on in-house data. 
At the bottom, we show the performance of the teacher model (consisting of 5 Bi-LSTM layers) used for distillation .}
\label{t:iemocap}
\begin{tabular}{lrr}
\hline
& \caja{c}{c}{Dev \\TER} & 
\caja{c}{c}{Test \\ WER} \\
\hline
% Train on Target (5 Uni-LSTM layers) & 74.00 & \\
Train on Target (4 Uni-LSTM layers) & 43.00 & \\
\hline
Our method (5 Uni-LSTM layers) &&\\
% \hline
Pretrain & 25.59 & 27.68 \\
$\quad$ + Finetune & 22.55& \\
$\qquad$ + LIN & 20.90 &  \\
$\qquad \quad$ + DataAug & 20.14 &  \\
$\qquad \qquad$ + KD on unsup (60 hours) & \textbf{18.75} & \textbf{20.64}\\
\hline
Teacher (5 Bi-LSTM layers, Pretrain&& \\
+ Finetune + LIN + DataAug) & 14.33 &   \\
$\qquad \qquad \qquad \qquad$ + word decode & 13.02 & \\
\hline
\end{tabular}
\end{table}

To show that this target dataset is much harder than WSJ, we train a model with 5 Uni-LSTM layers and another model with 4 Uni-LSTM layers from scratch on the target training data (18 hours). Due to the more challenging acoustic conditions and speaking style, both models perform very poorly, achieving the token error rate (TER) of 74\% and 43\% respectively. 
On the other hand, directly applying the uni-directional model pretrained on source domain data (15000 hours) to the target domain achieves a TER of 25.59\% on the dev set.
We then apply our methods to a model consisting of 5 Uni-LSTM layers. 
In Table~\ref{t:iemocap}, we show the results after finetuning, domain adaptation, data augmentation, and knowledge distillation on 60 hours of non-transcribed data. Consistent with results on WSJ, each technique leads to improvements in TER and their gains are additive. 
We measure the WER on test set for the pretrained model and final model, and observe more than 25.4\% relative improvement (27.68\%$\rightarrow$20.64\%) from our methods. % Note a bi-directional system is not deployable for an online system.

\section{Conclusions}
\label{s:conclusions}

We have shown that the careful combination of three data techniques significantly improves the ASR performance on two datasets of different speaking styles. Among these techniques, the use of pseudo-labels for distilling the modeling power from a domain-adapted bi-direction system to an uni-direction system with non-transcribed data is novel for our problem setup, is proven to be particularly effective, and shall be of interest to practitioners. In the future, we will explore these techniques with more advanced end-to-end methods.